\documentclass[submission, Phys]{SciPost}

\begin{document}

\begin{center}{\Large \textbf{
Compact bulk-machined electromagnets for quantum gas experiments
}}\end{center}

\begin{center}
K. Roux, B. Cilenti, V. Helson, H. Konishi, and J.P. Brantut\textsuperscript{1$*\dagger$}
\end{center}

\begin{center}
{\bf 1} Institute of Physics, École Polytechnique F\'ed\'erale de Lausanne, 1015 Lausanne, Switzerland
\linebreak
$^*$ jean-philippe.brantut@epfl.ch \\
$^\dagger$\href{http://lqg.epfl.ch}{http://lqg.epfl.ch}
\end{center}

\begin{center}
\today
\end{center}

\section*{Abstract}
{\bf
We present an electromagnet combining a large number of windings in a constrained volume with efficient cooling. It is based on bulk copper where a small pitch spiral is cut out and impregnated with epoxy, forming an ensemble which is then machined at will to maximize the use of the available volume. Water cooling is achieved in parallel by direct contact between coolant and the copper windings. A pair of such coils produces magnetic fields suitable for exploiting the broad Feshbach resonance of $^6$Li at $832.2\,$G. It offers a compact and cost-effective alternative solution for quantum gas experiments. }

\vspace{10pt}
\noindent\rule{\textwidth}{1pt}

\section{Introduction}
\label{sec:intro}

Strong, homogeneous magnetic fields of the order of a thousand Gauss are at the core of quantum gas experiments, from laser cooling and trapping \cite{Metcalf:2003aa} to the use of Feshbach resonances controlling the inter-atomic interactions \cite{Chin:2010aa}. As experimental apparatus become more and more complex, with more laser beams or high-aperture imaging systems to accommodate, space occupation as well as heat management constraints become more and more acute, calling for optimized and flexible electromagnet designs. Efforts in this direction have been reported in the past years, in particular novel designs for Zeeman slowers \cite{Bell:2010aa,Cheiney:2011aa}, Bitter electromagnets \cite{Bitter:1936aa,Bitter:1962aa,Sabulsky:2013ab} or improved heat management methods \cite{Ricci:2013aa}. More compact systems of magnetic traps use in-vacuum electromagnets \cite{Wang:2007aa,Saint:2018aa,Zhou:2017aa}, or atom chips \cite{Folman:2002aa,Reichel:2002aa}, but those are not adapted to experiments requiring large homogeneous fields. 

The common ground of these electromagnet concepts is the improvement over the widely used design based on wound copper wire (see for example \cite{Lewandowski:2003aa}). This solution is privileged due to its robustness, and the use of hollow wire allows for water cooling with good contact between copper and the coolant. It suffers however from several drawbacks, the first being the need for coolant and electrical current to follow the same path, yielding a large coolant pressure drop across the magnet even for a moderate number of turns. Several circuits of coolant fluid are then needed to limit temperature inhomogeneities in the coil. Another drawback is the lack of flexibility and poor space occupation efficiency: hollow copper wires are typically several millimeters wide, restricting the number of windings in a given volume. In addition, the epoxy matrix and coolant circuit inlets and outlets limit the fraction of volume actually used for current carrying copper. 

The Bitter-type configuration addresses some of these drawbacks \cite{Bitter:1936aa,Bitter:1962aa,Sabulsky:2013ab}, in particular it offers very good heat management thanks to the improved flow of coolant through the magnet, at the cost of requiring each winding to bear several coolant connections. Space occupation was addressed in a recent improvement where several layers are used to increase the number of windings \cite{Long:2018ab}. Other alternatives include the use of mixed configurations where hollow copper is combined with bulk copper wires \cite{Ganger:2018aa} in a single assembly, or a scheme with fully parallel cooling demonstrated in a few-windings coil directly machined from a bulk copper block \cite{Ricci:2013aa}. 

In this article, we present electromagnets optimizing space occupation, offering very large shape flexibility while allowing to reach the high fields required for the use of Feshbach resonances in $^6$Li \cite{Zurn:2013aa}. Our design is based on a bulk copper plate in which a spiral is cut by wire erosion \cite{Ricci:2013aa}, and impregnated with epoxy. The resulting ensemble is machinable using standard tools, allowing for carving out ridges to maximize space occupation or holes for electrical connections and clamping. As a result, the limited volume of a reentrant vacuum viewport accommodates $31$ windings over a single $22\,$mm thick layer. We present the detailed fabrication procedure of the ensemble, as well as performance in terms of magnetic fields and heat management, demonstrating the suitability of this approach for quantum gas experiments. 

\begin{figure}
\begin{center}
  \includegraphics[width=1\linewidth]{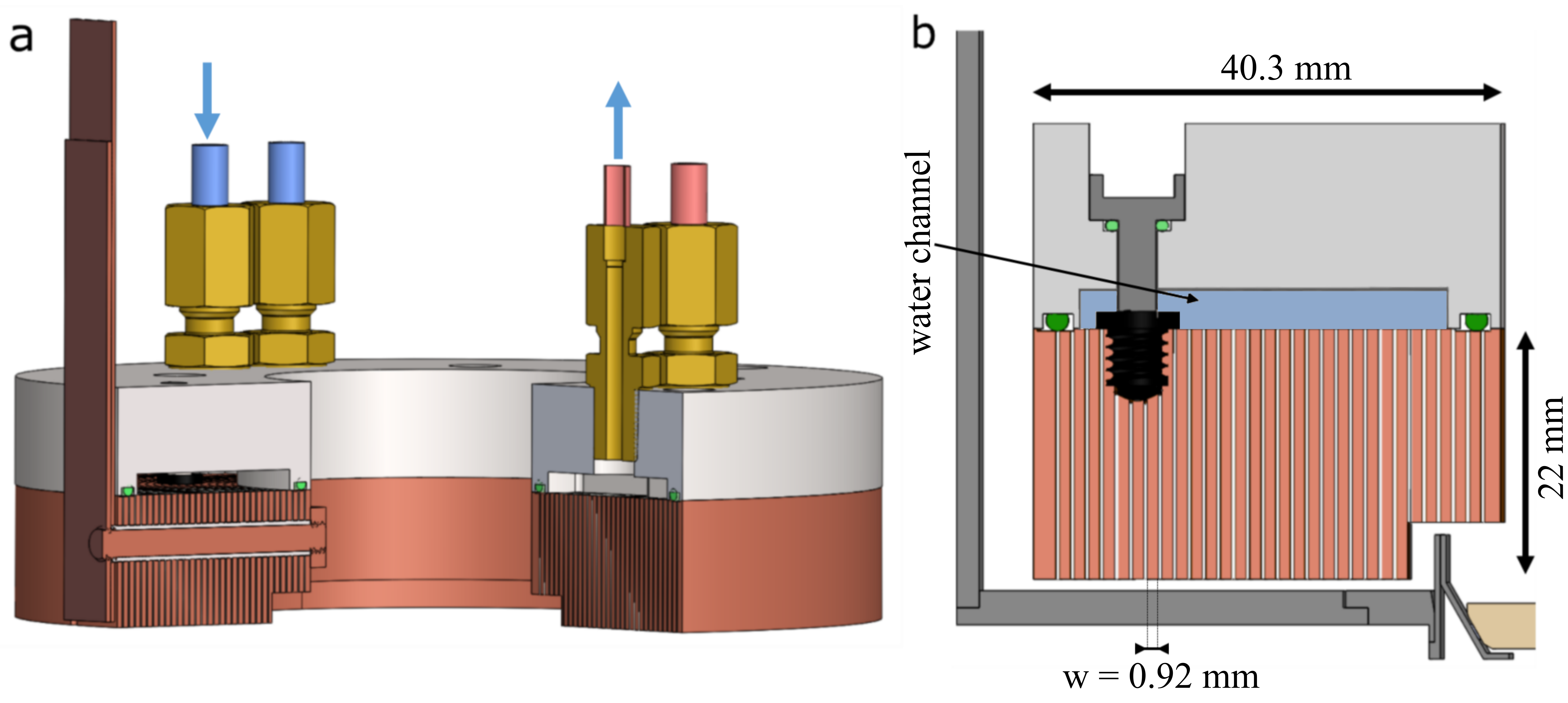}
  \caption{{\bf a}: 3D Computer-assisted drawing of the electromagnet. The copper coil consists of horizontally stacked windings (orange), separated by epoxy layers. A PEEK\textsuperscript{\ref{peek}} cap (light grey) creates a channel on top of the coil, rendered water tight by EPDM\textsuperscript{\ref{epdm}} o-rings (green), where cooling fluid is injected (light-blue arrows) and circulates in direct contact with the copper. At the bottom of the coil, a ridge is machined to fit the geometry of the reentrant viewport. A through hole allows for the electrical connection of the inner winding from the outside through the coil body. Electric current is injected and collected by a pair of flat cables running upward. {\bf b}: Cut view of the electromagnet, showing blind threaded holes with PEEK inserts (black) hosting titanium screws to hold the cap and the coil together. The resulting assembly optimally fits the available space in the reentrant viewport with current carrying copper.}
  \label{fig:fig1}
\end{center}
\end{figure}

\section{Concept and design}
\label{sec:concept}

The maximum magnetic field that a coil electromagnet can produce is limited by the geometric constraints on the coil dimensions and the total allowable power dissipation. In designing electromagnets for cold atoms experiments, the dimensions are constrained by the vacuum systems and necessary optical access, which limits the distance to the atoms and restricts the total volume available for the current carrying coil. In addition, the electric current is limited by practical constraints such as the diameter of cables, heating at the contacts and availability of power supplies. 

In our design, the electromagnet has to fit in the limited volume of a reentrant vacuum viewport, while allowing for the use of high-resolution optics.  At the same time, a pair of such electromagnets shall achieve magnetic fields of hundreds of Gauss several centimeters below the coil, in order to reach Feshbach resonances in $^6$Li atoms, with a current limited to $440$\,A by the power supply. We meet these requirements using a coil comprising $31$ turns horizontally stacked in a single $22\,$mm thick layer, depicted in figure \ref{fig:fig1}, with a large aspect ratio of $23.9$ for an individual winding. The inner and outer radii are $32$ and $72\,$mm respectively. 

With the geometry fixed, the performance is limited by power dissipation, which is achieved by forced convection with cold water. Three different processes determine cooling efficiency and thus the temperature distribution within the electromagnet: (i) heat conduction within the bulk of the coil, measured by the temperature difference across the copper coil $\Delta T_c$ (ii) heat transfer at the coil-coolant interface, measured by the temperature difference between the coil and the coolant $\Delta T_w$ and (iii) coolant flux through the assembly, measured by the coolant temperature difference $\Delta T_f$ between the inlet and outlet. The regime $\Delta T_f \gg \Delta T_c+ \Delta T_w$ corresponds to {\it flux-limited} cooling, where the coolant and magnet are thermalized and heat removal is limited by the flow, while $\Delta T_f \ll \Delta T_c+ \Delta T_w$ corresponds to a {\it transfer-limited} regime where heat transfer between the coil and the coolant limits the cooling. 

To guide our design, we estimate the parameter dependence of these three temperature differences for our horizontal stack configuration, where the electromagnet is cooled by cold water in direct contact with the upper surface of the coil. For a given current $I$ in the windings, neglecting lateral heat flow between windings, the temperature difference between the top and bottom of the coil reads
\begin{equation}
\Delta T_c = \frac{I^2}{w^2} \frac{\rho_\mathrm{Cu}}{2\lambda_\mathrm{Cu}},
\label{eq:Eq1}
\end{equation}
with $\rho_\mathrm{Cu}$ and $\lambda_\mathrm{Cu}$ the electrical resistivity and heat conductivity of copper, respectively, and $w$ the width of one winding (see Appendix A for the derivation). For a current of $400\,$A and $w=1\,$mm, we get $\Delta T_c = 3.4\,$K, showing that gradients within the coil will be minimal in spite of the very large aspect ratio for individual windings.

The heat transfer at the copper-water interface depends on the nature of the flow. For a typical total flow rate in the coil $0.23$ l$\cdot$s$^{-1}$, a simple estimate based on the hydraulic diameter of the duct yields turbulent flow with a Reynolds number of $\sim6.4\cdot10^3$\cite{Taine:2014aa}\footnote{The large aspect ratio and short length of the duct would call for a more detailed analysis of the flow, beyond the scope of the present work}. We deduce a heat transfer coefficient $h_w \sim5\cdot10^3\,\mathrm{W\cdot m^{-2} K^{-1}}$ and thus a temperature difference (see Appendix A for derivation)
\begin{equation}
\Delta T_w = I^2 \frac{\rho_\mathrm{Cu} }{H w^2 h_w}.
\label{eq:Eq2}
\end{equation}
with $H$ the coil thickness. For our design, we evaluate the ratio $\Delta T_c / \Delta T_w \sim 0.14$, which indicates that the interface is the limiting factor in the cooling \footnote{Within the fluid, heat is carried predominantly by convection rather than conduction, as indicated by a Nusselt number of $\sim 70$\cite{Taine:2014aa}}. 

Last, we can easily estimate the increase of water temperature across the coil from simple energy conservation considerations, yielding 
\begin{equation}
\Delta T_f = I^2 \frac{R}{C_wQ},
\end{equation}
where $R$ is the total electrical resistance of the coil, $Q$ is the coolant flux and $C_w$ is the volumetric heat capacity of water. For our system allowing large water fluxes of $0.23$ l$\cdot$s$^{-1}$, we estimate $\Delta T_f \sim 0.7\,$K for $I=400\,$A. 

%In this configuration, convection is indeed the dominant heat transfer process, both within the fluid, where we estimate a Nusselt number of the order of $70$, and between the coil and the coolant: to assess the relative roles of heat diffusion within the copper and heat exchange at the interface, we form the dimensionless ratio 
%\begin{equation}
%\frac{h_w H }{2 \lambda_\mathrm{Cu}} =\frac{\Delta T}{\Delta T_w}, 
%\label{eq:Eq2}
%\end{equation}
%with $H$ the thickness of the coil, giving the ratio between the temperature differences across the copper coil $\Delta T$ and between the water and copper $\Delta T_w$, which we expect to be $\sim0.2$. Interestingly, this does not depend on the interface area.

This shows that even with an extreme aspect ratio for the coil windings, heat diffuses efficiently through the coil up to the water-copper interface, and that we operate well in the transfer-limited regime in contrast to most designs \cite{Lewandowski:2003aa,Sabulsky:2013ab}. Operating in the transfer-limited regime has the advantage that heat removal is less sensitive to the water flux since the flux enters only through the heat transfer coefficient with a sub-linear dependence \cite{Taine:2014aa}.

\section{Manufacturing and assembly }
\label{sec:manufacturing}

Manufacturing of the electromagnet followed the concept of \cite{Ricci:2013aa}, where a coil is carved out from bulk copper rather than wound out of wires. As described below, a major advantage of this concept is that the thick body of the coils forms a single, rigid ensemble, which can then be shaped using standard lathe and milling machines. As long as machining does not significantly affect the windings, the final shape can fulfill a wide variety of space constraints, with negligible effects on the magnetic field distribution. We used this capability in order to carve out tapped holes for holding the cap, a through hole carrying current through the coil body from the inner to the outer part and an edge in order to fit the exact geometry of the reentrant viewport, as can be seen in figure \ref{fig:fig1}.

\subsection{Coil body}

Manufacturing started from a $24\,$mm thick oxygen-free copper plate. Wire-erosion machining creates a spiral with a pitch of $1.3\,$mm, leaving a gap between consecutive windings of $0.38\,$mm. Fiber-glass reinforced plastic spacers were inserted between successive windings to preserve uniform spacing and ensure electrical insulation. After cleaning and drying, the spiral was impregnated with epoxy glue. We used commercial low viscosity, room-temperature curing epoxy\footnote{Sicomin SR1710/SD7820}, loaded with $10\%$ of $0.25\,\mu$m fiber glass flake and $30\%$ aluminium nitride (AlN) powder\footnote{Sigma Aldritch 10$\mu$m powder, $\geq 98 \%$ purity}. Preliminary tests showed that cracks can develop in the inter-windings regions due to thermal stress when using high temperature curing expoxy, hence the use of a fiber glass load to reinforce the structure. We also found that the use of AlN load improves the heat dissipation from the inner and outer windings of the coils, which are not in direct contact with water. 

The impregnated coil was then machined with a lathe in order to remove the excess glue and expose the copper on the coil facet to be water cooled. The resulting surface quality is high, as can be seen in figure \ref{fig:figPhoto}, demonstrating the efficiency of machining of the composite assembly. Further machining could be used to roughen the surface on purpose, favoring turbulent flow to improve heat transfer. The assembly consists in more than $70\%$ copper in volume. Blind threaded holes were machined on the top surface to fit screws pressing the cap against the coil. A transverse through-hole was drilled and an fiber-glass insulating tube glued inside to accommodate a $5.5$ mm diameter copper screw carrying current from the inner to outer winding (see figure \ref{fig:fig1}).

\subsection{Electromagnet assembly}

\begin{figure}[htb]
\begin{center}
  \includegraphics[width=0.75\linewidth]{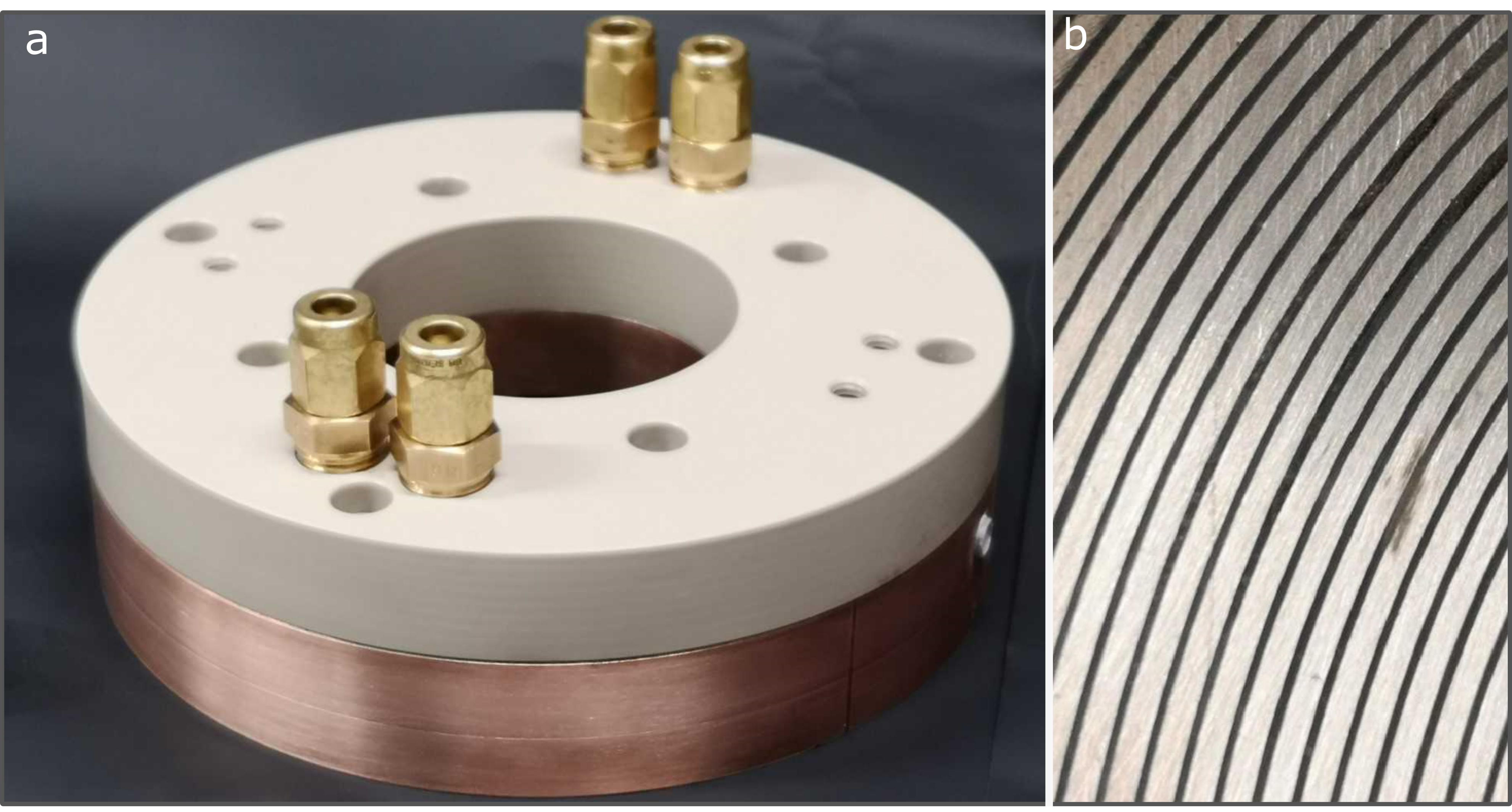}
  \caption{{\bf a}: Photograph of the assembled electromagnet, showing the PEEK cap attached to the coil body, with coolant connections from the top. {\bf b}: Close view of the top surface of the coil after glueing and machining on a standard lathe, showing the successive windings separated by the electrically insulating epoxy. The width of a copper winding is $0.92\,$mm.}
  \label{fig:figPhoto}
\end{center}
\end{figure}

As shown in figure \ref{fig:fig1}, the electromagnet mainly consists of the coil and the plastic cap creating the water cooling channel and holding the ensemble together. The cap is a U-shaped PEEK\footnote{Polyether ether ketone\label{peek}} part forming a $3\,$mm high channel on the top surface of the coil, with a cross-sectional area of $102$\,mm$^{2}$. To increase the water flux, the channel is split in two half-circles with separate inlets. The cap is attached to the coil body by $8$ titanium screws fitting in threaded holes machined on the top surface of the coil, with peak inserts for electrical isolation. Titanium offers high stiffness, is non-magnetic, and is more cathodic than copper which reduces the risk of galvanic corrosion. PEEK inserts fit into the threaded holes to electrically insulate the coil from the screws, and holes are rendered leak tight using EPDM\footnote{Ethylene Propylene Diene Methylene rubber. EPDM 70, Shore A, 1.5\,mm cord, 6\,mm diameter \label{epdm}} O-rings. Two long O-rings pressed in the inner and outer sides of the cap ensure that the ensemble is leak-tight\footnote{We use Curil T (Elring) grease for matching the O-rings with the surface.} up to 4 bars overpressure. 

Electric current is injected and collected via two copper plates separated by a thin electrical insulation, minimizing stray magnetic fields (see figure \ref{fig:fig1}). Current is collected from the inner side via a copper screw going through the coil itself. This configuration, made possible by our machinable coil concept, further optimizes the use of space and minimizes stray fields since it does not require an electrical wire running from the inner part of the coil\footnote{We use Chemtronics CW7100 grease for contact between the copper bolt and the windings}. 

The resulting system, shown in figure \ref{fig:figPhoto}, is highly compact. The cap is used to suspend the ensemble such that all the available space of the reentrant viewport is occupied by current carrying copper, maintained as close as possible to the vacuum chamber surface but mechanically disconnected from it.

\section{Performance}
\label{sec:performance}
\subsection{Electromagnetic properties}
\begin{figure}[htb]
\begin{center}
  \includegraphics[width=1\linewidth]{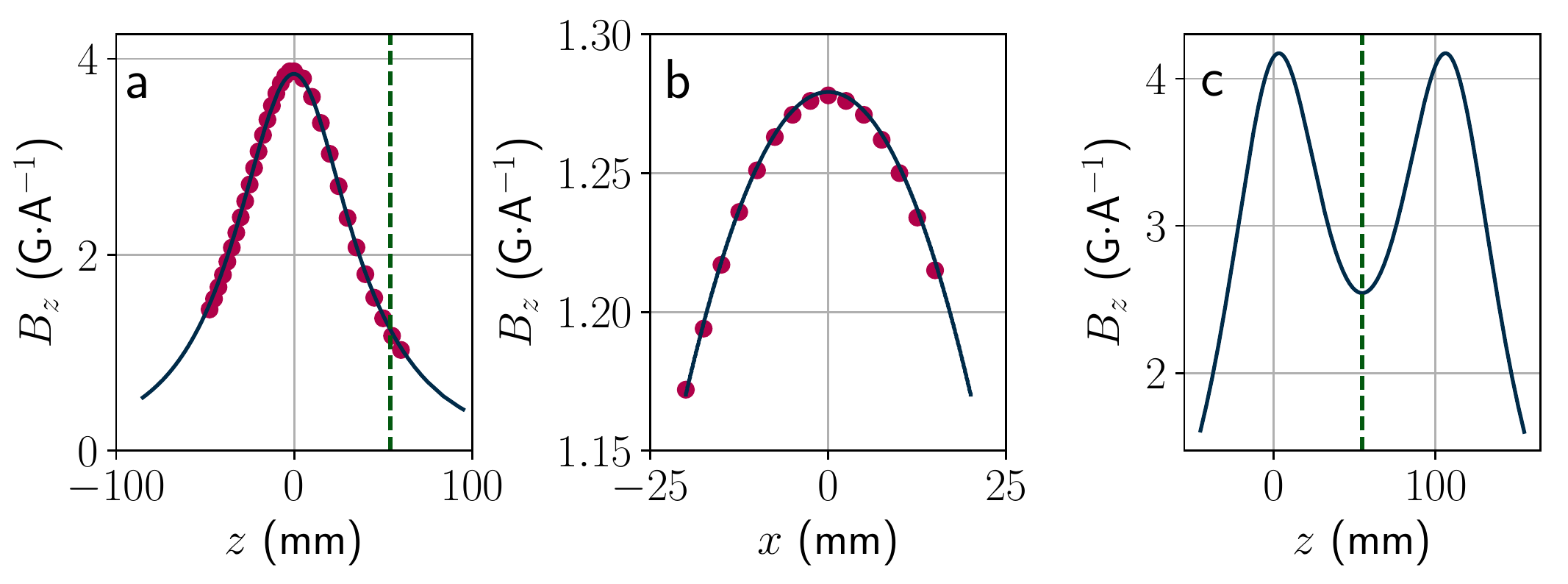}
  \caption{Axial component of the magnetic field created by one electromagnet as a function of {\bf a}: the position $z$ along the coil axis, and {\bf b}: the position $x$ in the plane of the atoms located $52.2$ mm from the center of the coil (dashed vertical line on {\bf a}). The red points are measurements, and the solid lines represent simulations using the Radia package.  {\bf c}: Simulated axial magnetic field created by a pair of identical coils separated by $104.4$\,mm. The dashed vertical line represents the expected position of the atoms.}
  \label{fig:figMag}
\end{center}
\end{figure}

We first characterized the magnetic field distribution with a DC current and the electromagnet in the steady state. The axial component of the magnetic field $B_z$ produced by the electromagnet was measured with a uniaxial Hall probe\footnote{Lakeshore 425 with HMNA-1904-VR probe} as a function of position. The results are presented in figure \ref{fig:figMag}. For comparison, we performed numerical simulations using the Radia package \cite{Chubar:1998aa} considering homogeneous current distributions inside the coil body. These are presented as solid lines in the figure, showing very good agreement with the measurements. 

We measured a peak magnetic field at the expected position of the atoms, $52.2$\,mm below the mean coil position, of $1.28\,\mathrm{G}\cdot\mathrm{A}^{-1}$, implying that a pair of such coils separated by $104.4$\,mm will yield a field of $832.2$\,G, the position of the broad Feshbach resonance of $^6$Li \cite{Zurn:2013aa}, at a current of $325$\,A. Our simulation predicts the field distribution for this pair of coils, as shown in figure \ref{fig:figMag}c. At the position of the atoms, the field varies quadratically with a curvature along the axial direction of $0.11$\,$\mathrm{G}\cdot\mathrm{A}^{-1}\mathrm{cm}^{-2}$. The combination of a finite curvature, yielding magnetic trapping of high-field seeking atoms along the radial directions, with a dipole trap providing confinement along the axial direction is a standard configuration for cold $^6$Li atoms experiments \cite{Ketterle:2008aa}. Variations of magnetic fields across a typical atomic cloud close to the Feshbach resonance are negligible compared with the $262\,$G width of the resonance. The simulation also showed that the current carried in the through hole from the inner to outer winding, which breaks the cylindrical symmetry of the ensemble, results in negligible disturbance in the field distribution.

We measured the inductance of the coil, which characterizes its dynamical response. Its impedance was measured as a function of frequency between $0.16\,$Hz and $320\,$Hz. We fitted a DC resistance of $10.4(10)\,$m$\Omega$ and an inductance of $116(2)$\,$\mu$H, in reasonable agreement with the magnetic field simulation yielding $94$\,$\mu$H and with the expected intrinsic resistance of $8.7\,$m$\Omega$. This is comparable with the results obtained for Bitter electromagnets with similar geometry and number of turns \cite{Sabulsky:2013ab}. 

\subsection{Heat management}

\begin{figure}[t]
\begin{center}
  \includegraphics[width=1\textwidth]{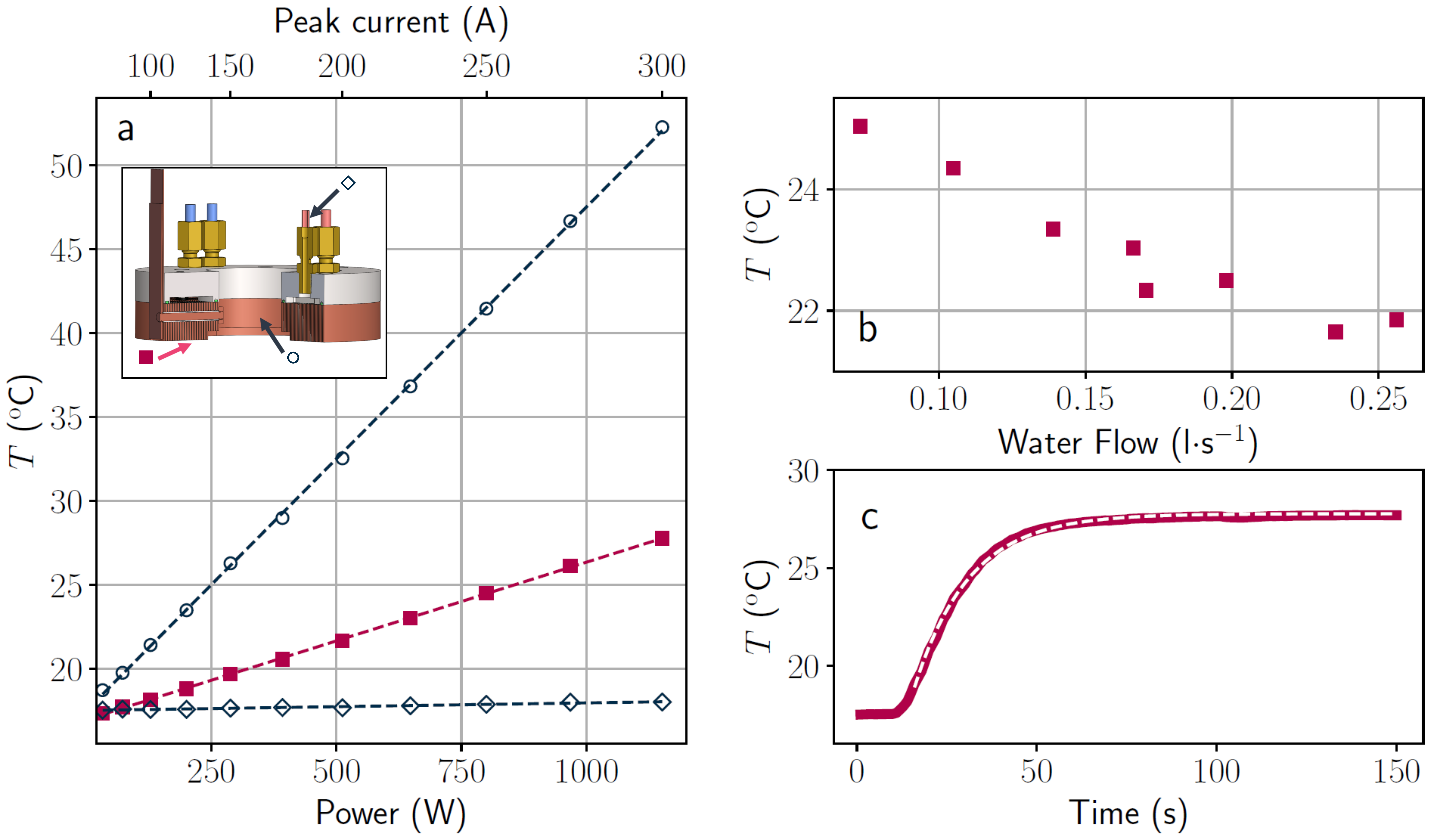}
  \caption{{\bf a}: Steady state temperature of the electromagnet as a function of electrical power dissipated with a total coolant flux of $0.23$ l$\cdot$s$^{-1}$, measured at the inner winding (circles), and bottom surface (squares), and temperature of the water exiting the magnet (diamonds). The dashed lines are linear fit to the data, yielding heating rates of $0.45$, $9.4$ and $30$\,K$\cdot$kW$^{-1}$ for the water, coil body and inner winding, respectively. {\bf b}: Temperature of the coil body as a function of the total water flux in the coil, for an average electrical power of $512$\,W. {\bf c}: Time evolution of the temperature of the coil body following a sudden switch-on of $300$\,A. The thick red line represents measurements and the dashed line is an exponential fit yielding a timescale of $15.26\,$s.}
  \label{fig:figTh}
\end{center}
\end{figure}

We tested the thermal performance of the electromagnet by measuring the temperature at different points of the assembly during operation. We first imposed a fixed cooling water flux of $0.23$ l$\cdot$s$^{-1}$ with a water temperature of $17.5^{\mathrm{o}}$C. The steady state temperature as a function of current is presented in figure \ref{fig:figTh}a at various locations in the system. The overall temperature of the coil body, except for the first inner and outer winding, measured on the bottom side of the coil, remains below $30^{\mathrm{o}}$C. This confirms our expectations that heat is efficiently removed by water running over the edges of the coil windings. The heating rate of the coil body is $9.4$\,K$\cdot$kW$^{-1}$.

The water temperature increase across the system was very low, and remained lower than at any other point in the electromagnet, confirming that it operates in the transfer-limited regime. We observed that the temperature is larger at the inner and outer windings than in the bulk of the ensemble, because these windings are not in direct contact with water, as can be observed in figure \ref{fig:fig1}. This results from the design choice to privilege the number of windings over the thermal homogeneity\footnote{Using the same machining capabilities, our concept allows for the use of a variable pitch spiral ensuring direct water cooling of each winding. In the present case, this would have reduced the number of windings by $\sim 4$}. Operating the electromagnet in realistic experimental conditions, with $350\,$A current and a duty cycle of $30\%$, we observed that the highest temperature was reached in the inner winding, with a steady-state value of $34^{\mathrm{o}}$C, compatible with routine operations in the laboratory.

We then measured the temperature variations with coolant flux. Thanks to the wide section of the coolant circuit, the water flux was only limited by the diameter of the inlet pipes, allowing for total fluxes up to $0.26\,\mathrm{l}\cdot\mathrm{s}^{-1}$ for a moderate water pressure drop of $3.5$ bars across the electromagnet. The heating of the coil body is reduced upon increasing the water flux, which is expected due to the dependence of the heat transfer coefficient on velocity in the turbulent regime. 

Finally, we measured the dynamics of the thermalization of the coils, monitoring the temperature after switching on the electrical current to $300$\,A, with a water flux of $0.23$ l$\cdot$s$^{-1}$. The result is shown in figure \ref{fig:figTh}c. The evolution is well fitted by a single exponential, which timescale $\tau$ allows for an estimate of the average copper-water heat transfer coefficient $h$: energy balance considerations, supposing a homogeneous temperature within the coil, yields $\tau \sim wC_\mathrm{Cu}/h =15.3$\,s, with $C_\mathrm{Cu}$ the specific heat of copper, from which we obtain $h \sim 5\cdot10^3$\,W$\cdot$m$^{-2}$K$^{-1}$, in agreement with our initial expectations. 

This short thermalization time is favorable since thermal expansion of the coils changes the magnetic field, such that accurate calibration requires a steady-state situation. For our geometry, we estimate that changes of the coil radius with temperature produce relative variations of the magnetic fields of $\sim 1.3\cdot10^{-5} \,\mathrm{K}^{-1}$. Thermal effects upon varying the coil temperature will be comparable with the effects of finite accuracy of standard power supplies.

\section{Conclusions}
\label{sec:conclusion}

We have presented a compact and flexible electromagnet concept adapted to quantum gas experiments requiring both large homogeneous magnetic fields and minimal space occupation. The overall costs are minimal, limited to that of the raw copper, inexpensive epoxy glue, and a home made custom plastic mold for the gluing operation. Cutting the spiral required $\sim40$ hours on the wire erosion machine of our institute's mechanical workshop, running fully automatically.  

Our experimental system is constrained by the need for high numerical aperture together with the accommodation of a large, in-vacuum experimental platform, leaving only limited space for the main electromagnets. Thanks to the compactness of the main electromagnets, the reentrant viewports of our experimental setup can also accommodate 4 pairs of compensation coils in the cloverleaf configuration \cite{Mewes:1996aa}, allowing for moving the saddle point of the field in the plane of the coil. This possibility will allow for precise positioning of the atomic cloud with respect to the experimental platform. It will also allow for transport measurements with lithium quantum gases in the two-terminal configuration \cite{Krinner:2017aa}.

\section*{Acknowledgements}
We acknowledge the technical assistance of Claude Amendola, Olivier Haldimann, Gilles Grandjean, Philippe Zuercher and Damien Fasel. 

\paragraph{Funding information}
We acknowledge funding from the ERC project DECCA (Project No. 714309), the Sandoz Family Foundation-Monique de Meuron program for Academic Promotion and EPFL. 

\begin{appendix}

\section{Temperature gradients across the electromagnet}
\subsection{Conduction through the coil body}

\begin{figure}[htb]
\begin{center}
  \includegraphics[width=0.3\textwidth]{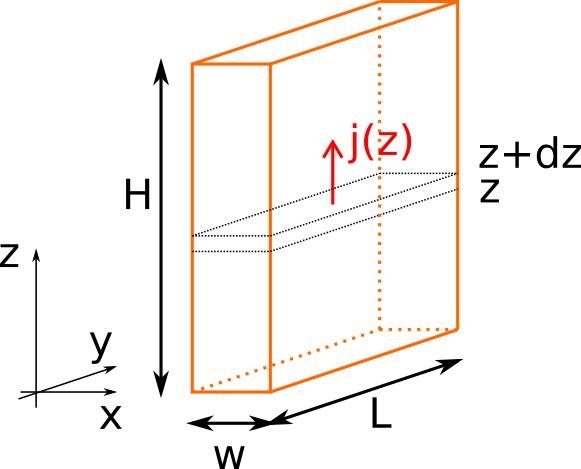}
  \caption{Calculation of the temperature profile within one winding, based on energy balance within a slice of thickness $dz$.}
  \label{fig:Eq1}
\end{center}
\end{figure}

We consider one winding of the coil as a prism with rectangular section of width $w$ and height $H$, with a total length $L$, as illustrated in figure \ref{fig:Eq1}. With the notations of the main text, the total power dissipated is 
\begin{equation}
P = \rho_\mathrm{Cu} \frac{L}{Hw} I^2 = p LHw
\end{equation}
where we introduced the power dissipated per unit volume $p$. 

To calculate the temperature profile we neglect heat flowing on the sides, and consider the temperature as homogeneous in the $x-y$ plane. In the steady state, the energy contained in the slice between $z$ and $z+dz$ is constant, imposing  
\begin{equation}
\frac{dj}{dz} = p 
\end{equation}
where $j$ is the heat flux through unit area. With $j(0) = 0$ we get $j(z) = pz$. Fourier's law then relates the temperature profile $T(z)$ to the heat flux
\begin{equation}
\lambda_\mathrm{Cu} \frac{dT}{dz} = j(z)
\end{equation}
which yields 
\begin{equation}
T(z) = T(0) + \frac{pz^2}{2\lambda_\mathrm{Cu}  } = T(0) + I^2 \frac{\rho_\mathrm{Cu} }{2\lambda_\mathrm{Cu}  w^2} \frac{z^2}{H^2}
\end{equation}
Evaluating this expression for $z=H$ yields equation \ref{eq:Eq1}. The dependence on $H$ has dropped since upon varying the thickness at fixed total current, the increase of heat resistance is balanced by the decrease of total dissipated power.

\subsection{Coil-water interface}

We consider the temperature at the coil-coolant interface in the presence of forced convection. In steady state, the continuity of the heat flux at the interface imposes that 
\begin{equation}
h_w(T(H) - T_w) = Hp
\end{equation}
where $T_w$ and $T(H)$  are the temperatures of the coolant and coil at height $H$ calculated above, $h_w$ is the heat transfer coefficient at the interface, and $p$ is the power dissipation per unit volume within the coil. We obtain  
\begin{equation}
\Delta T_w = I^2 \frac{\rho_\mathrm{Cu} }{H w^2 h_w}.
\end{equation}

%We evaluate the relative contributions to cooling of the heat transfer at the interface and heat conduction within the coil, we compare the temperature differences between the top and bottom of the coil $\Delta T$ with the temperature difference between the coolant and coil surface $\Delta T_w = T(H) - T_w$ yielding equation \ref{eq:Eq2}.

\end{appendix}

\bibliography{paper_coil}

\begin{thebibliography}{10}
\providecommand{\url}[1]{\texttt{#1}}
\providecommand{\urlprefix}{URL }
\expandafter\ifx\csname urlstyle\endcsname\relax
  \providecommand{\doi}[1]{doi:\discretionary{}{}{}#1}\else
  \providecommand{\doi}{doi:\discretionary{}{}{}\begingroup
  \urlstyle{rm}\Url}\fi
\providecommand{\eprint}[2][]{\url{#2}}

\bibitem{Metcalf:2003aa}
H.~J. {Metcalf} and P.~{van der Straten},
\newblock \emph{{Laser cooling and trapping of atoms}},
\newblock Journal of the Optical Society of America B Optical Physics
  \textbf{20}, 887 (2003),
\newblock \doi{10.1364/JOSAB.20.000887}.

\bibitem{Chin:2010aa}
C.~Chin, R.~Grimm, P.~Julienne and E.~Tiesinga,
\newblock \emph{Feshbach resonances in ultracold gases},
\newblock Rev. Mod. Phys. \textbf{82}, 1225 (2010),
\newblock \doi{10.1103/RevModPhys.82.1225}.

\bibitem{Bell:2010aa}
S.~C. Bell, M.~Junker, M.~Jasperse, L.~D. Turner, Y.-J. Lin, I.~B. Spielman and
  R.~E. Scholten,
\newblock \emph{A slow atom source using a collimated effusive oven and a
  single-layer variable pitch coil zeeman slower},
\newblock Review of Scientific Instruments \textbf{81}(1), 013105 (2010),
\newblock \doi{10.1063/1.3276712}.

\bibitem{Cheiney:2011aa}
P.~Cheiney, O.~Carraz, D.~Bartoszek-Bober, S.~Faure, F.~Vermersch, C.~M. Fabre,
  G.~L. Gattobigio, T.~Lahaye, D.~Gu{\'e}ry-Odelin and R.~Mathevet,
\newblock \emph{A zeeman slower design with permanent magnets in a halbach
  configuration},
\newblock Review of Scientific Instruments \textbf{82}(6), 063115 (2011),
\newblock \doi{10.1063/1.3600897}.

\bibitem{Bitter:1936aa}
F.~Bitter,
\newblock \emph{The design of powerful electromagnets part ii. the magnetizing
  coil},
\newblock Review of Scientific Instruments \textbf{7}(12), 482 (1936),
\newblock \doi{10.1063/1.1752068}.

\bibitem{Bitter:1962aa}
F.~Bitter,
\newblock \emph{Water cooled magnets},
\newblock Review of Scientific Instruments \textbf{33}(3), 342 (1962),
\newblock \doi{10.1063/1.1717838}.

\bibitem{Sabulsky:2013ab}
D.~O. Sabulsky, C.~V. Parker, N.~D. Gemelke and C.~Chin,
\newblock \emph{Efficient continuous-duty bitter-type electromagnets for cold
  atom experiments},
\newblock Review of Scientific Instruments \textbf{84}(10), 104706 (2013),
\newblock \doi{10.1063/1.4826498}.

\bibitem{Ricci:2013aa}
L.~Ricci, L.~M. Martini, M.~Franchi and A.~Bertoldi,
\newblock \emph{A current-carrying coil design with improved liquid cooling
  arrangement},
\newblock Review of Scientific Instruments \textbf{84}(6), 065115 (2013),
\newblock \doi{10.1063/1.4811666}.

\bibitem{Wang:2007aa}
R.~Wang, M.~Liu, F.~Minardi and M.~Kasevich,
\newblock \emph{Reaching 7li quantum degeneracy with a minitrap},
\newblock Phys. Rev. A \textbf{75}, 013610 (2007),
\newblock \doi{10.1103/PhysRevA.75.013610}.

\bibitem{Saint:2018aa}
R.~Saint, W.~Evans, Y.~Zhou, T.~Barrett, T.~M. Fromhold, E.~Saleh, I.~Maskery,
  C.~Tuck, R.~Wildman, F.~Oru{\v c}evi{\'c} and P.~Kr{\"u}ger,
\newblock \emph{3d-printed components for quantum devices},
\newblock Scientific Reports \textbf{8}(1), 8368 (2018),
\newblock \doi{10.1038/s41598-018-26455-9}.

\bibitem{Zhou:2017aa}
Y.~{Zhou}, N.~{Welch}, R.~{Crawford}, F.~{Oru{\v c}evi{\'c}}, F.~{Wang},
  P.~{Kr{\"u}ger}, R.~{Wildman}, C.~{Tuck} and T.~M. {Fromhold},
\newblock \emph{{Design of Magneto-Optical Traps for Additive Manufacture by 3D
  Printing}},
\newblock ArXiv e-prints  (2017),
\newblock \eprint{1704.00430}.

\bibitem{Folman:2002aa}
R.~{Folman}, P.~{Kr{\"u}ger}, J.~{Schmiedmayer}, J.~{Denschlag} and
  C.~{Henkel},
\newblock \emph{{Microscopic Atomic Optics: From Wires to an Atomic Chip}},
\newblock Advances in Atomic and Molecular Physics \textbf{48}, 263 (2002),
\newblock \eprint{0805.2613}.

\bibitem{Reichel:2002aa}
J.~{Reichel},
\newblock \emph{{Microchip traps and Bose-Einstein condensation}},
\newblock Applied Physics B: Lasers and Optics \textbf{74}, 469 (2002),
\newblock \doi{10.1007/s003400200861}.

\bibitem{Lewandowski:2003aa}
H.~J. Lewandowski, D.~M. Harber, D.~L. Whitaker and E.~A. Cornell,
\newblock \emph{Simplified system for creating a bose--einstein condensate},
\newblock Journal of Low Temperature Physics \textbf{132}(5), 309 (2003),
\newblock \doi{10.1023/A:1024800600621}.

\bibitem{Long:2018ab}
Y.~Long, F.~Xiong, V.~Gaire, C.~Caligan and C.~V. Parker,
\newblock \emph{All-optical production of $^{6}\mathrm{Li}$ molecular
  bose-einstein condensates in excited hyperfine levels},
\newblock Phys. Rev. A \textbf{98}, 043626 (2018),
\newblock \doi{10.1103/PhysRevA.98.043626}.

\bibitem{Ganger:2018aa}
B.~G{\"a}nger, J.~Phieler, B.~Nagler and A.~Widera,
\newblock \emph{A versatile apparatus for fermionic lithium quantum gases based
  on an interference-filter laser system},
\newblock Review of Scientific Instruments \textbf{89}(9), 093105 (2018),
\newblock \doi{10.1063/1.5045827}.

\bibitem{Zurn:2013aa}
G.~Z\"urn, T.~Lompe, A.~N. Wenz, S.~Jochim, P.~S. Julienne and J.~M. Hutson,
\newblock \emph{Precise characterization of li6 feshbach resonances using
  trap-sideband-resolved rf spectroscopy of weakly bound molecules},
\newblock Phys. Rev. Lett. \textbf{110}, 135301 (2013),
\newblock \doi{10.1103/PhysRevLett.110.135301}.

\bibitem{Taine:2014aa}
J.~Taine, F.~Enguehard and E.~Lacona,
\newblock \emph{Transferts thermiques: Introduction aux transferts
  d'{\'e}nergie},
\newblock Dunod,
\newblock ISBN 9782100714582 (2014).

\bibitem{Chubar:1998aa}
O.~Chubar, P.~Elleaume and J.~Chavanne,
\newblock \emph{{A three-dimensional magnetostatics computer code for insertion
  devices}},
\newblock Journal of Synchrotron Radiation \textbf{5}(3), 481 (1998),
\newblock \doi{10.1107/S0909049597013502}.

\bibitem{Ketterle:2008aa}
W.~{Ketterle} and M.~W. {Zwierlein},
\newblock \emph{{Making, probing and understanding ultracold Fermi gases}},
\newblock Nuovo Cimento Rivista Serie \textbf{31}, 247 (2008),
\newblock \doi{10.1393/ncr/i2008-10033-1},
\newblock \eprint{0801.2500}.

\bibitem{Mewes:1996aa}
M.-O. Mewes, M.~R. Andrews, N.~J. van Druten, D.~M. Kurn, D.~S. Durfee and
  W.~Ketterle,
\newblock \emph{Bose-einstein condensation in a tightly confining dc magnetic
  trap},
\newblock Phys. Rev. Lett. \textbf{77}(3), 416 (1996),
\newblock \doi{10.1103/PhysRevLett.77.416}.

\bibitem{Krinner:2017aa}
S.~Krinner, T.~Esslinger and J.-P. Brantut,
\newblock \emph{Two-terminal transport measurements with cold atoms},
\newblock Journal of Physics: Condensed Matter \textbf{29}(34), 343003 (2017),
\newblock \doi{10.1088/1361-648X/aa74a1}.

\end{thebibliography}

\nolinenumbers

\end{document}